\documentclass{PoS}
\newcommand{\be}{\begin{equation}}
\newcommand{\ee}{\end{equation}}
\newcommand{\beqn}{\begin{eqnarray}}
\newcommand{\eeqn}{\end{eqnarray}}
\newcommand{\eq}[1]{(\ref{#1})}

\newcommand{\dd}{{\mathrm d}}

\newcommand{\Tr}{{\mathrm{Tr}\,}}
\newcommand{\Z}{{\mathbb{Z}}}
\newcommand{\Dirac}{\rlap {\hspace{-0.01mm} \slash} D}
\newcommand{\dirac}{\rlap {\hspace{-0.5mm} \slash} \partial}

\title{Phase diagram of strong interactions in an external magnetic field}

\ShortTitle{Phase diagram of hot QCD in an external magnetic field}

\author{\speaker{Ana J\'ulia ~Mizher}\\
        
Centro Brasileiro de Pesquisas Fisicas, Rua Xavier Sigaud 150, 22290-180\\
 Rio de Janeiro, RJ, Brazil\\
        E-mail: \email{amizher@cbpf.br}}

\author{{Eduardo S. ~Fraga}\\       
Instituto de F\'\i sica, Universidade Federal do Rio de Janeiro, \\
Caixa Postal 68528, Rio de Janeiro, RJ 21941-972, Brazil.\\
        E-mail: \email{fraga@if.ufrj.br}}
        
\author{{M. N. ~Chernodub}\thanks{On leave from ITEP, Moscow, Russia.}\\
CNRS, Laboratoire de Math\'ematiques et Physique Th\'eorique, Universit\'e Fran\c{c}ois-Rabelais,
F\'ed\'eration Denis Poisson - CNRS, Parc de Grandmont, Universit\'e de Tours, 37200 France
Department of Physics and Astronomy, Ghent University, Krijgslaan 281, S9, B-9000 Gent,
Belgium\\
E-mail: \email{maxim.chernodub@lmpt.univ-tours.fr}}

\abstract{We obtain the phase diagram of strong interactions in the presence of a magnetic field 
within the linear sigma model coupled to quarks and to the Polyakov loop, and show that the chiral 
and deconfinement lines can split. We also study the behavior of the chiral condensate in this 
magnetic environment and find an approximately linear dependence on the external field, 
in accordance with lattice data.} 

\FullConference{The many faces of QCD\\
		November 2-5, 2010\\
		Gent Belgium}

\begin{document}

It is believed that experiments with ultrarelativistic heavy ion collisions can provide access to some features of the vacuum of QCD. In this regime the QCD coupling constant is very strong, making any investigation, experimentally and theoretically, very involved. Perturbation theory does not apply in this case and non-perturbative methods indicate the existence of non-trivial topological solutions, accessible only via semi-classical treatments. It was proposed in Ref. \cite{Kharzeev:2007jp} that if such solutions are indeed present in the strongly interacting matter generated after a heavy ion collision, its association with the strong magnetic field produced in non-central collisions can induce a phenomenon that has been called ``chiral magnetic effect'' (CME). This 
effect generates a current of the deconfined quarks that causes a separation of electric
charges in the final hadronized state. Some suitable observables can be defined in order to analyze 
the charge separation caused by the CME~\cite{Voloshin:2000xf}.

In this work we investigate the effects of a magnetic background on the QCD phase transitions \cite{Fraga:2008qn,ref:main} using the linear sigma model (LSM) coupled to quarks and to the Polyakov loop (PLSMq). The chiral and deconfinement transitions are temperature driven and a finite-temperature mean-field approach is implemented in order to model this scenario. 

The confining properties are described by the Polyakov loop $L$, whose expectation value works as an order parameter that indicates whether the system resides in the confined or deconfined phase:
\beqn
\mbox{Confinement}:\quad
\left\{
\begin{array}{llll}
\langle L \rangle  & = & 0 \quad , \quad & \mbox{low $T$}   \\
\langle L \rangle  & \neq & 0 \quad , \quad & \mbox{high $T$}
\end{array}
\right.\,, \qquad\quad L(x) = \frac{1}{3} \Tr {\cal P} \exp \Bigl[i
\int\limits_0^{1/T} \dd \tau \, A_4(\vec x, \tau) \Bigr]\,,
\label{eq:L} \label{eq:L:phases} \eeqn
where $A_4$ is the matrix-valued temporal component of the Euclidean
gauge field $A_\mu$ and the symbol ${\cal P}$ denotes path ordering. The
integration takes place over the compactified imaginary time~$\tau$.

On the other hand, the chiral sector of the theory is described by the 
LSM part of the model, whose degrees of freedom are the $\sigma$ and $\pi$ mesons. The expectation value of the $\sigma$ meson field works as an order parameter for the chiral symmetry, indicating if this symmetry is broken or restored, depending on the temperature:
\beqn
\mbox{Chiral symmetry}:\quad
\left\{
\begin{array}{llll}
\langle \sigma \rangle  & \neq & 0 \quad , \quad & \mbox{low $T$}   \\
\langle \sigma \rangle  & = & 0 \quad ,\quad & \mbox{high $T$}
\end{array}
\right.\,,
\label{eq:sigma:phases}
\qquad\qquad
\begin{array}{lll}
\phi & = & (\sigma,\vec{\pi})\,,\\
\vec{\pi} & = & (\pi^{+},\pi^{0},\pi^{-})\,.
\end{array}
\label{eq:phi}
\eeqn
The spontaneous and explicit chiral symmetry breaking features are encoded in the LSM potential:
\beqn
V_\phi(\sigma ,\vec{\pi}) = \frac{\lambda}{4}(\sigma^{2}+\vec{\pi}^{2} - {\it v}^2)^2-h\sigma\, .
\eeqn
Hot quarks provide  a background in which the condensates evolve. 
The interaction of the quarks with the mesons is implemented in a Yukawa fashion, with the following Lagrangian:
\be
{\cal L}_q =
\overline{\psi}_f \left[i\gamma ^{\mu}\partial _{\mu} - g(\sigma +i\gamma _{5}
\vec{\tau} \cdot \vec{\pi} )\right]\psi_f \,,
\ee
where $g$ is the coupling constant. 

The static uniform magnetic field,
pointing in the $z$ direction, is introduced 
by a replacement of the simple derivative by a covariant derivative in the quark kinetic term. This derivative contains an Abelian gauge field which encodes all the relevant information about the magnetic field. In the same way, the interaction with the Polyakov loop is introduced via a non-Abelian gauge field in the covariant derivative as follows:
\be
\Dirac = \gamma^{\mu} (\partial _{\mu} - i Q\, a_\mu - i A_\mu),
\label{eq:covariant}
\ee
and the full Lagrangian can now be written as:
\beqn
{\cal L} =  \overline{\psi} \left[i \Dirac - g(\sigma +i\gamma _{5}
 \vec{\tau} \cdot \vec{\pi} )\right]\psi + \frac{1}{2}\bigl[(\partial _{\mu}\sigma)^2 + (\partial _{\mu} \pi^0)^2\bigr]
 + |D^{(\pi)}_\mu \pi^+|^2  - V_\phi(\sigma ,\vec{\pi}) - V_L(L,T)\,,
\label{eq:L:full}
\eeqn
where $D_\mu^{(\pi)} = \partial_\mu + i e a_\mu$ is the covariant derivative acting on colorless charged pions, $a_{\mu}$ representing the $4$-potential for the external magnetic field, $V_\phi(\sigma ,\vec{\pi})$ is the LSM potential and $V_L(L,T)$ the Polyakov loop potential, given by \cite{ref:ratti08}:
\beqn
\frac{V_L(L,T)}{T^4} =-\frac{L^*L}{2}\sum_{l=0}^2 a_l \left(\frac{T_0}{T}\right)^l + b_3\left(\frac{T_0}{T} \right)^3\,\ln\left[1-6\,L^*L+4\left({L^*}^3+L^3\right) - 3\left(L^*L\right)^2\right]\,,
\label{eq:VL}
\eeqn
with $T_0 \equiv T_{SU(3)} = 270\, \mbox{MeV}$ being the critical temperature
in the pure $SU(3)$ gauge theory and
$a_0 = 16 \pi^2/45 \approx 3.51$, $a_1 = -2.47$, $a_2 = 15.2$, and $b_3 = -1.75$
are the phenomenological parameters.

The one-loop corrections  to the free energy $\Omega$ coming from quarks can be written as:
\beqn
e^{i V_{3d} \, \Omega_q/T}  =
\left[\frac{\det(i \Dirac^{(q)} - m_q )}{\det (i \dirac - m_q)} \right] 
 \cdot
\left[\frac{\det_T(i \Dirac^{(q)} - m_q)}{\det (i \Dirac^{(q)} - m_q )}\right]\,.
\label{eq:Omega:quark2}
\eeqn
The integration measure in the presence of the magnetic field $B$ for $T=0$ and $T>0$ becomes, respectively, as follows:
\begin{eqnarray}
&&\int \frac{d^4k}{(2\pi)^4} \mapsto \frac{|q B|}{2\pi}\sum_{n=0}^\infty 
\int \frac{dk_0}{2\pi}\frac{dk_z}{2\pi} \, ,\\
T\sum_{\ell} &&\int \frac{d^3k}{(2\pi)^3} \mapsto \frac{|q B| T}{2\pi}\sum_\ell \sum_{n=0}^\infty 
\int \frac{dk_z}{2\pi} \,.
\end{eqnarray}
The presence of the magnetic field generates a sum over the Landau levels $n$ 
in addition to the standard finite-temperature sum over the Matsubara frequencies 
\beqn
\omega_\ell = 2 \pi T \ell\,,
\eeqn
labeled by the integer number $\ell$.

The expectation values can be obtained minimizing the free energy,
\beqn
\Omega(\sigma,L;T,B) & = & V_\phi(\sigma,\vec\pi) + V_{L}(L,T)
+ \Omega_q(\sigma,L,T) \,, 
\label{eq:Omega}
\eeqn
at fixed values of temperature and magnetic field.
The interaction piece $\Omega_q(\sigma,L,T)$ can be separated in two parts: a vacuum part 
-- which is temperature independent but still magnetic-field dependent -- 
and a thermal piece. Up to one loop the vacuum part is given by:
\beqn \nonumber
\Omega^{\mathrm{vac}}_{q}(B) & = &
\frac{1}{i V_{4d}} \log \left[\frac{\det(i \Dirac^{(q)} - m_q )}
{\det (i \dirac - m_q)} \right]
\label{eq:Omega:q:vac}\\ \nonumber
& & \hskip -15mm\\
&=&-\frac{N_{c}}{\pi}
\sum_{f=u,d}|q_{f}|B \left[ \left(
\sum_{n=0}^\infty 
I_{B}^{(1)}(M_{nf}^{2})\right)
\right. 
 - \left.
\frac{I_{B}^{(1)}(m_{f})}{2}\right] \,
\eeqn
minus the  standard vacuum correction in the absence of the magnetic field,
\beqn
\Omega_{q}^{(0)}&=& 2N_{c}\sum_{f=u,d}I_{B}^{(3)}(m_{f}^{2}) \, ,
\eeqn
where we have defined the integral
\beqn
I_{B}^{(d)}(M^{2})&=& \int \frac{d^{d}p}{(2\pi)^{d}} \sqrt{p^{2}+M^{2}}.
\eeqn
Here capital $M$ means an effective mass which takes into account 
the effect of the magnetic field $B$:
\beqn
M^2_{nf} = m^2_f + 2 n |q_f B|\,.
\eeqn

The thermal piece is given by:
\beqn
\Omega^{\mathrm{para}}_q(\sigma,\Phi) & = &
\frac{T}{i V_{3d}} \ln \left[\frac{\det_T(i \Dirac^{(q)} - m_q)}{\det (i \Dirac^{(q)} - m_q )}\right]
\nonumber\\
& = &  - \frac{|q| B T}{2\pi} \sum_{i=1}^3 \sum_{s = \pm \frac{1}{2}}
\sum_{n=0}^\infty \sum_{\ell\in\Z}
\int\limits_{-\infty}^{+\infty}
\frac{d p_z}{2 \pi} \ln \Bigl[\Bigl(\frac{\omega_\ell}{T}
+ \frac{A^{ii}_4}{T}\Bigr)^2 + \frac{\omega^2_{sn}(p_z,\sigma)}{T^2} \Bigr]\,,
\label{eq:Vpsi:1}
\eeqn
where the dispersion relation for quarks in the external magnetic field is
\beqn
\omega_{sn}(p_z,\sigma) = {\bigl[m_q^2(\sigma) + p_z^2 + (2 n + 1 - 2 s) |q| B\bigr]}^{1/2}\,,
\label{eq:omega:sn}
\eeqn
with constituent quark mass $m_q(\sigma)= g \sigma$. The thermal part~\eq{eq:Vpsi:1} includes interaction of the Polyakov loop with quarks and therefore it breaks the $\Z_3$ symmetry, which was initially present in the potential for the Polyakov loop $V_L$.

The vacuum correction has remarkable consequences on the phase structure. In previous work involving the LSM, 
this correction was usually not taken into account.
However, in a theory with spontaneous symmetry breaking, the presence of a condensate will 
modify the masses, so that contributions from vacuum diagrams 
can not be subtracted away as trivial zero-point energies. These contributions were shown 
to play an important role at finite temperature by the authors of Ref. \cite{Mocsy:2004ab}, who 
incorporate scale effects phenomenologically. Vacuum contributions were considered 
at finite density in the perturbative massive Yukawa model with analytic exact results up to 
two loops in Refs. \cite{Palhares:2008yq,thesis} and, more specifically, in optimized perturbation 
theory at finite temperature and chemical potential in Ref. \cite{Fraga:2009pi}. More recently, 
this issue was discussed in different scenarios for the chiral and deconfining transitions (see, 
e.g., Refs. \cite{ref:vacuum:logs}). The exclusion of the vacuum correction was 
motivated by the phenomenological argument that quark degrees of freedom are absent in the vacuum. On the other hand, models like the Nambu-Jona-Lasinio (NJL) include such vacuum corrections automatically. Which treatment represents a better 
description of QCD is still subject of debate, so that we opted for studying both cases. 

\begin{figure}[!thb]
\vskip 3mm
\begin{center}
\includegraphics[width=85mm,clip=true]{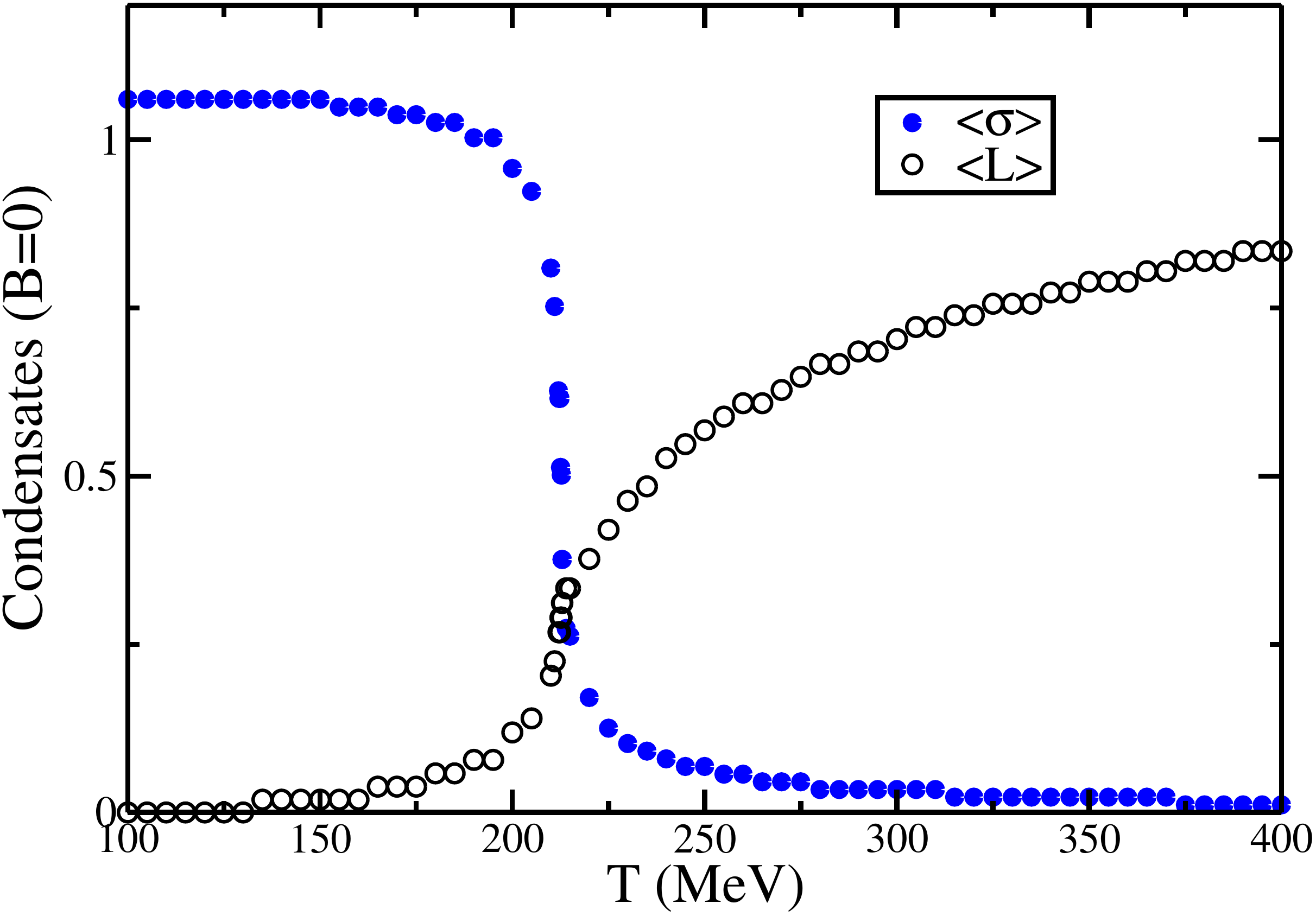}
\end{center}
\vskip -3mm
\caption{
Expectation values
of the order parameters for the chiral and deconfinement transitions as functions of the temperature. The filled circles represent the
$\sigma$--condensate and the empty circles stand for the expectation value of the Polyakov loop. Both lines are smooth functions of the temperature
indicating the presence of a crossover. In this plot the condensates are dimensionless.}
\label{fig:B0}
\end{figure}

For finite temperature and $B=0$ we obtain a crossover for both transitions, as shown in Fig.\ref{fig:B0}. We define the critical temperature as the one where the curve changes curvature, which happens simultaneously for the chiral and deconfinement transitions.

\begin{figure}[!thb]
\begin{center}
\begin{tabular}{cc}
\includegraphics[width=70mm,clip=true]{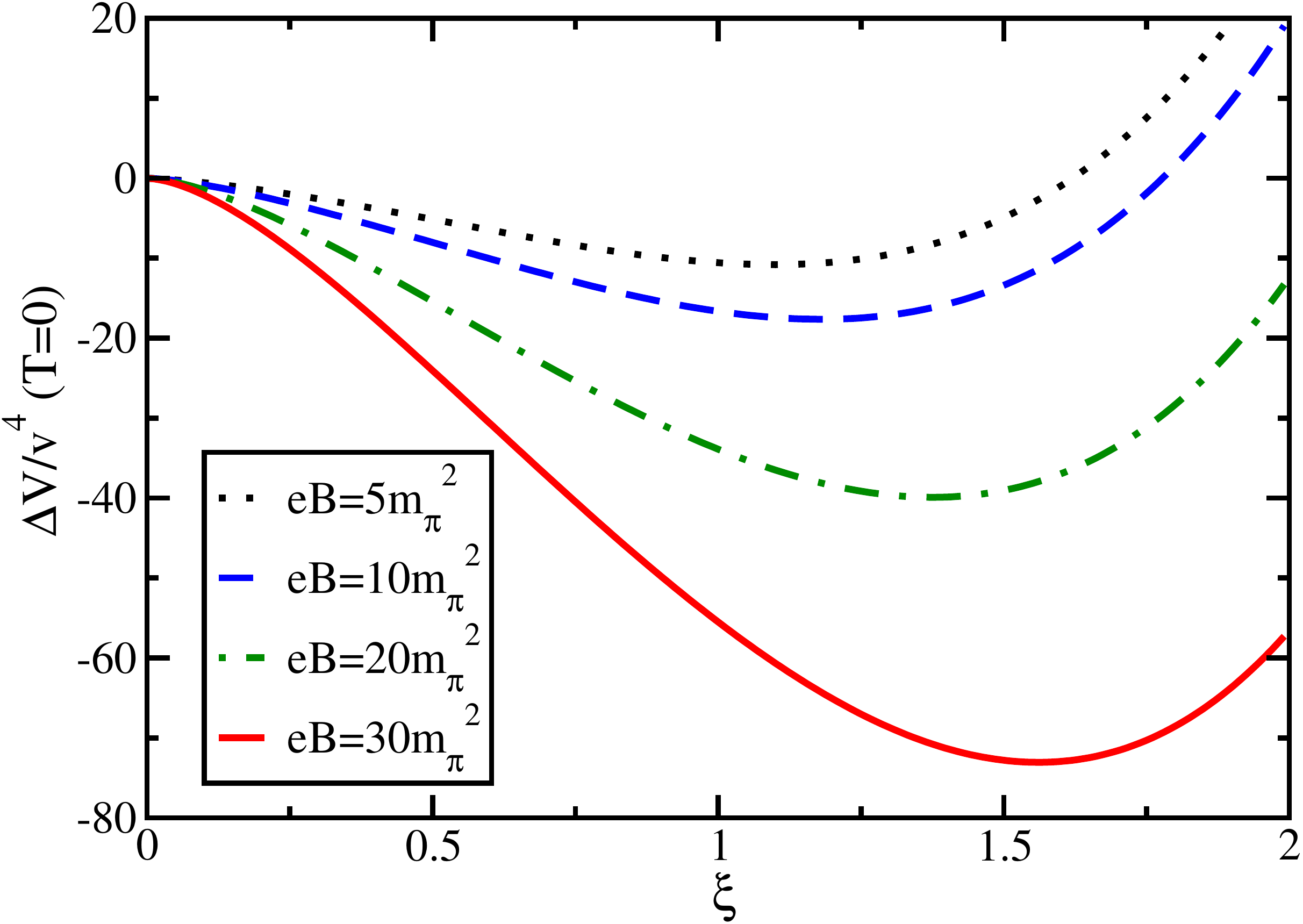} 
& \includegraphics[width=75mm,clip=true]{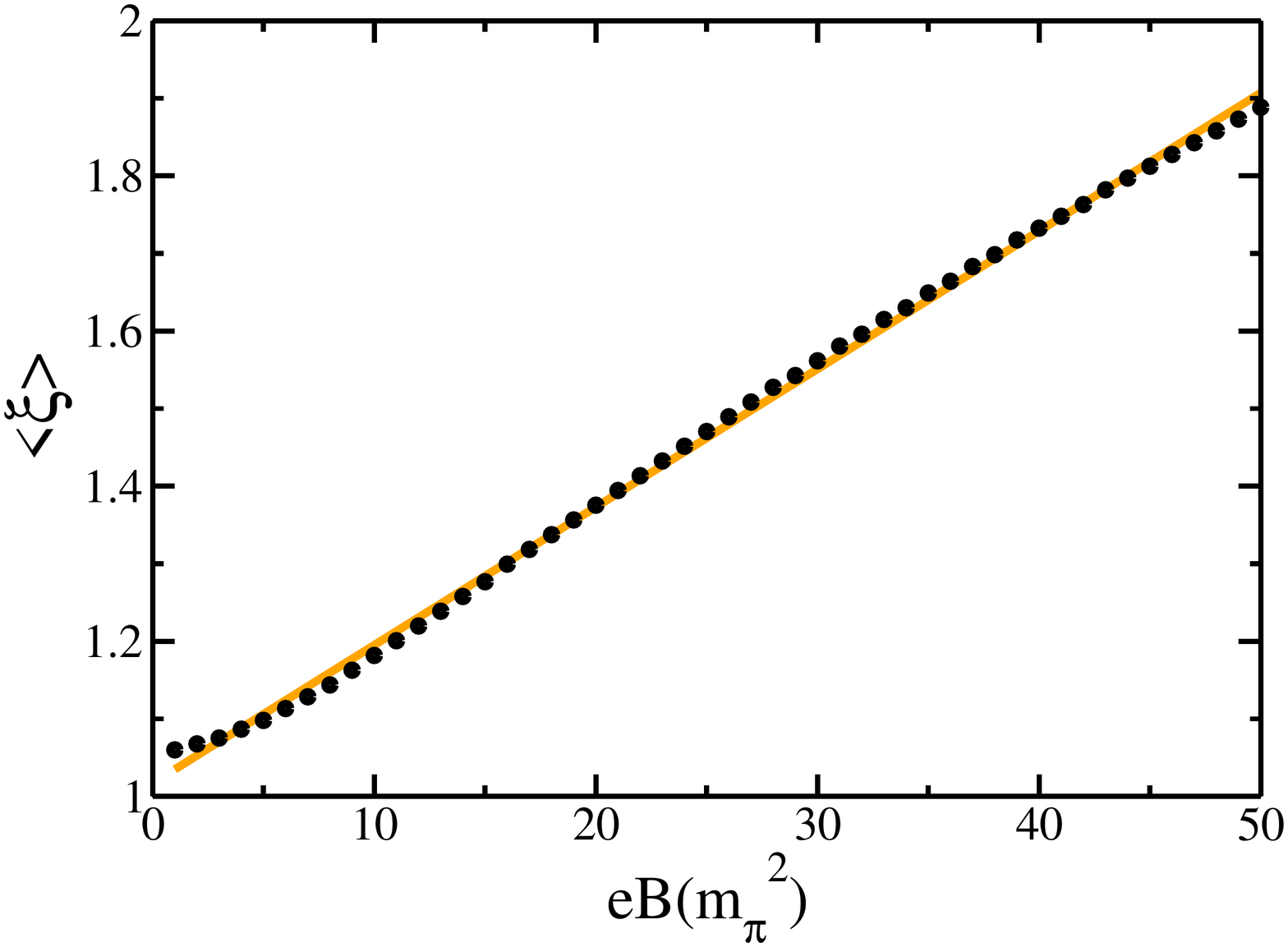}\\[-3mm]\\
\end{tabular}
\end{center}
\caption{Left: effective potential for the condensate $\sigma = \xi v$ at zero temperature for several values of the magnetic field $B$. Magnetic catalysis is manifested in the increase of the condensate and in the deepening of the potential. Right: expectation value of the condensate as a function of the magnetic field. The black dots are obtained from the PLSMq and the orange line is a linear fit~(19).}
\label{fig:ximin_B}
\end{figure}

At zero temperature, the presence of the magnetic field enhances the breaking of chiral symmetry, making the potential of the chiral field $\sigma$
deeper and increasing the value of the chiral condensate (in the zero-temperature case the Polyakov loop variable does not play a role). This enhancement effect is known as magnetic catalysis~\cite{ref:catalysis}, and is illustrated by the left plot of Fig. \ref{fig:ximin_B}, where the behavior of the zero-temperature potential for several values of the magnetic field is shown. 

\begin{figure}[!thb]
\begin{center}
\begin{tabular}{cc}
\qquad no vacuum corrections & \qquad with vacuum corrections \\
\includegraphics[width=70mm,clip=true]{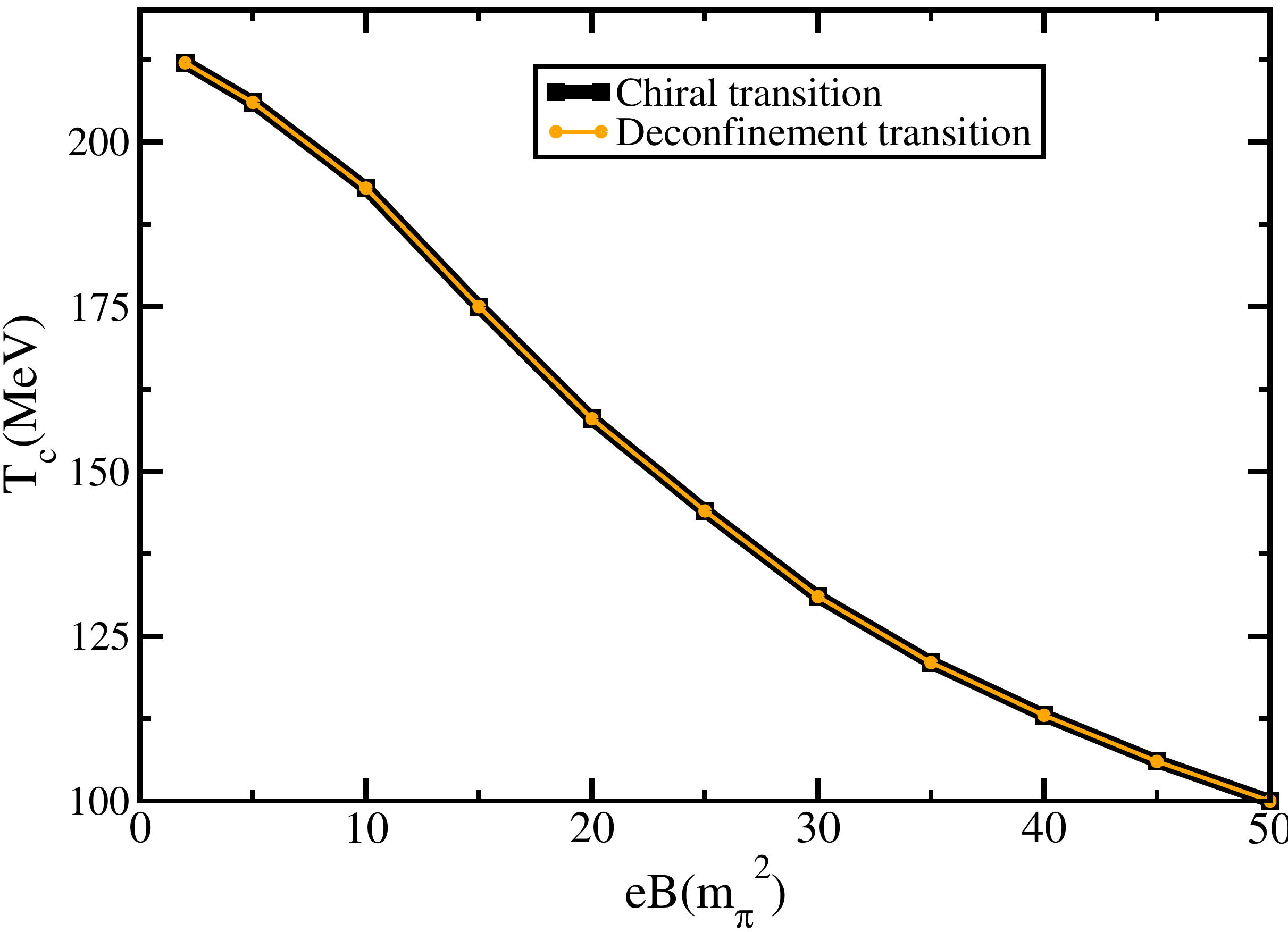} 
& \includegraphics[width=70mm,clip=true]{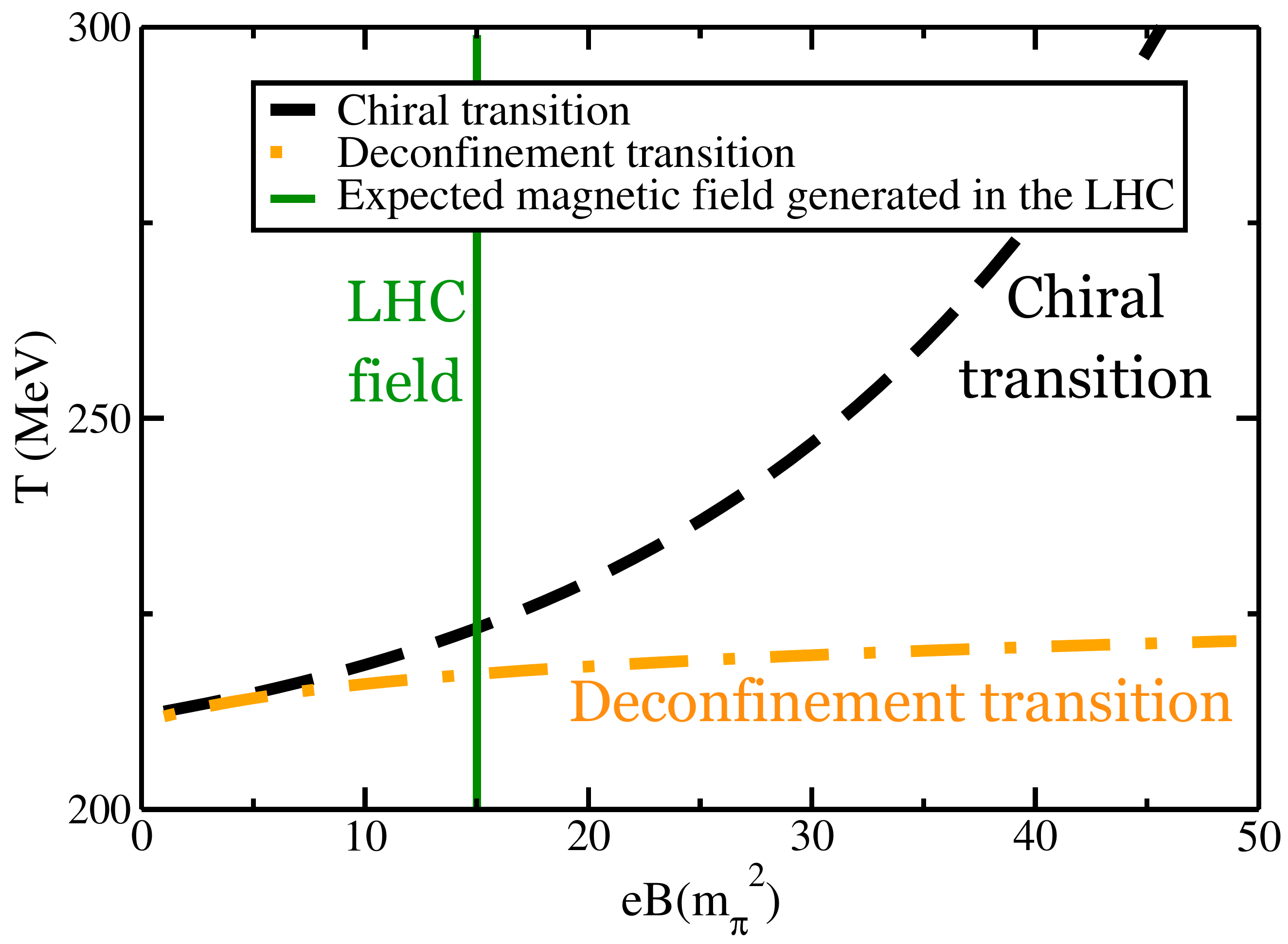}\\[-3mm]
\end{tabular}
\end{center}
\caption{Phase diagram in the $B$-$T$ plane.
Left: without $B$--dependent vacuum corrections. The critical temperatures of the deconfinement (dash-dotted line)
and chiral (dashed line)
transitions coincide all the way, and decrease with $B$.
Right: with $B$--dependent vacuum corrections: the critical temperatures of the deconfinement (dash-dotted line)
and chiral (dashed line) transitions 
coincide at $B=0$ and split
at higher values of the magnetic field. A deconfined phase with broken chiral symmetry appears.
The vertical line
represents a typical
magnitude of the magnetic field expected to be realized in heavy-ion collisions at the LHC ~\cite{Skokov:2009qp}.}
\label{fig:phase:diagram}
\end{figure}

It is interesting to study the dependence of the chiral condensate on the magnitude of the magnetic field. 
Analytic calculations in chiral perturbation theory~\cite{ref:linear} and results from lattice QCD~\cite{Buividovich:2008wf} indicate a linear dependence of the condensate on the strength of the magnetic field. Our model is consistent with this picture, and a linear curve fits well the results obtained within our approach, as can be seen in the right plot of Fig.~\ref{fig:phase:diagram}. The best linear fit in the plot is 
\beqn
\xi \equiv \frac{\sigma}{v} = 1.017 + 0.018 |eB|\,, \qquad v=87.73\,\mbox{MeV}\,.
\label{eq:fit}
\eeqn
The small deviation of the curve from the linear behavior is qualitatively similar to the one found in the context of the Sakai-Sugimoto holographic model \cite{Callebaut:2011uc}, although smaller.
For a small strength of the magnetic field, the dependence of the constituent quark mass (which plays the same role as the chiral condensate in our approach) 
on the magnetic field in Ref.~\cite{Callebaut:2011uc} shows a quadratic behavior, which can be observed in our plot as well.

For the full potential, including thermal and magnetic effects, we studied separately the cases with and without vacuum corrections. We obtained phase diagrams for both cases, showing that the effect of vacuum corrections is indeed remarkable. Fig.~\ref{fig:phase:diagram} shows the corresponding phase diagrams.

If the vacuum corrections are not included, the magnetic field turns both chiral and deconfinement transitions into first order transitions. Both transitions happen at the same critical temperature, which decreases as the magnetic field increases. If included, the vacuum corrections dominate the potential, changing drastically the picture. The transitions become crossovers and split as the magnetic field is increased. In both scenarios the presence of the magnetic field have important consequences, either turning the transitions into first order or causing a split between them, generating a new phase. It also reinforces the breaking of the $\Z_3$ symmetry that occurs when the Polyakov loop interacts with quarks. 

The scenario with vacuum corrections included, Fig.~\ref{fig:phase:diagram} (right), agrees well with a similar calculation performed in the NJL model \cite{Gatto:2010qs}, as was expected since the latter includes quark degrees of freedom in the vacuum by construction. In this case, the mentioned new phase -- where the quarks are already deconfined, but chiral symmetry is still broken -- should also occur. Recent calculations in a modified version of the NJL model suggest the existence of a quantitatively new, third, scenario where the gap between the two transitions is very small, while the nearly-common critical temperature increases with the magnetic field \cite{Gatto:2010pt}. Lattice simulations of QCD with dynamical quarks also indicate a rise of the critical temperatures and show no evidence of the splitting~\cite{D'Elia:2010nq}. However, in this lattice simulation the dynamical quarks are still heavier compared to realistic values. On the other hand, a holographic calculation points out to the existence of a small splitting between the transitions~\cite{Callebaut:2011uc}.

\acknowledgments
The authors thank David Dudal and Marco Ruggieri for useful discussions. This work was partially supported by CAPES-COFECUB project 663/10, CNPq, FAPERJ, FUJB/UFRJ and by the grant ANR-10-JCJC-0408 HYPERMAG.

\end{document}